\documentclass[%
 reprint,%
 amssymb, amsmath,%
 aip,jap,%
]{revtex4-1}

\usepackage{docs}%
\usepackage{graphicx}%
\usepackage{bm}%
\usepackage[colorlinks=true,linkcolor=blue]{hyperref}%
\expandafter\ifx\csname package@font\endcsname\relax\else
 \expandafter\expandafter
 \expandafter\usepackage
 \expandafter\expandafter
 \expandafter{\csname package@font\endcsname}%
\fi
\begin{document}

\title{First-principles study on electron field emission from nanostructures}%

\author{Hyon-Chol Choe}
\author{Nam-Hyok Kim}
\author{Hyok Kim}
\author{Song-Jin Im}
\email{ryongnam7@yahoo.com}
\affiliation{Department of Physics, Kim Il Sung University, Daesong District, Pyongyang, Democratic People's Republic of Korea}

\begin{abstract}
A first-principles approach is introduced to calculate electron field emission characteristics of nanostructures, based on the nonequilibrium Green function technique combined with the density functional theory. The method employs atomic-like basis orbitals with large confinement radii and lithium anode to describe the electron density in the vacuum between the nanostructure tip and the anode, so takes the presence of emitted current into account. The simulation results on a capped single-walled carbon nanotube reproduce the experimental trend closely, in particular, the current saturation and the deviation from the Fowler-Nordheim behavior.
\end{abstract}
\maketitle

Electron field emission (FE) has been one of active research areas owing to its theoretical, as well as commercial, significance.\cite
{1} In recent years, various nanostructures, including carbon nanotubes, metal oxide nanowires and nanoneedles, have been synthesized, and among many applications areas, they have emerged as promising candidates for electron emitters due to their very high aspect ratios.\cite
{2,3,4,5,6,7} Actually, it has been demonstrated that single-walled carbon nanotubes (SWNT) provide low turn-on voltages and brightness values of an order of magnitude higher than typical conductors such as copper and silver.\cite
{2} Given these experimental results, it is important to develop theoretical methods which can calculate and analyze FE characteristics of any nanostructures accurately.

The first model for FE from metal surfaces was proposed by Fowler and Nordheim in 1928, which assumes that under an external electric field electrons in the metal tunnel through a one-dimensional (1D) potential barrier into the vacuum.\cite
{8} The emitted current, when graphically expressed on ln[$IE$ $^{-2}$]
versus E$^{-1}$ scales, where $I$ is current and $E$ is electric field, exhibits a linear relationship, which is known as the Fowler-Nordheim (FN) plot. This model has been widely used by experimentalists to interpret the performance of electron emitters like the nanostructures.\cite
{2,3,4,5,6,7} However, the full applicability of the FN model to the FE from the nanostructures is not obvious: electron states in the nanostructures are not similar to that in the metal, and the nanostructures are not an infinitely wide surface, but three-dimensional (3D) structures. Thus, a more general and acurate treatment than the FN model is required, and among various approaches first-principles methods are the most adequate for this goal. 

There have been some first-principles methods to calculate the FE characteristics of the nanostructure emitters. Han and co-workers calculated the many-body wave function of the emitting nanotube using a pseudopotential-based time-dependent density functional theory approach.\cite
{9,10} However, in order to calculate transmission functions they used a 1D potential barrier. Next, Khazaei $et$  $al$. developed an approach to calculate the FE properties based on first-principles local density of states and effective potentials.\cite
{11} Although it has been used in several theoretical works on the FE from various nanostructures,\cite
{12,13,14,15} it also uses a 1D potential barrier and the WKB approximation to calculate the tunneling current. Recently, Yaghoobi and co-workers proposed a method based on a real-space first-principles Hamiltonian in the nonequilibrium Green's function (NEGF) and Fisher-Lee's transmission formulation, which took the 3D nature of the problem into account.\cite
{16} However, their first-principles Hamiltonian and NEGF are constructed in the absence of electronic current.

To the best of our knowledge, a first-principles approach based on the electron density and the Hamiltonian determined self-consistently in the presence of electronic current, has not been reported to date. Here, for the first time we present a first-principles method for calculating the FE characteristics of nanostructures using the NEGF technique combined with the density functional theory (DFT),\cite
{17,18,19} which allows the nonequilibrium conditions to be taken into account.

In FE experiments using a nanostructure as an electron emitter, the nanostructure is grown on a cathode electrode and placed in front of an anode at some distance. As it has been shown that most of the nanostructure remains equipotential under an external field and potential drop occurs mainly very close to the tip of nanostructure,\cite
{20} the FE characteristics of the nanostructure can be determined by the region that encapsulates the nanostructure tip and vacuum. Therefore, it is reasonable to simulate only a small section close to the tip of nanostructure, which allows us to study the FE using fully first-principles approaches. However, such a short section of the nanostructure would not reproduce the strong field enhancement that the entire nanostructure would have due to its high aspect ratio. Although in order to compensate for the lack it is common to use an electric field value that already contains the effect of field enhancement,\cite
{16} typically several hundred times stronger than the applied external field,\cite
{21} we will not introduce such parameters in the simulation. 

Our method employs the state-of-the-art quantum transport algorithms based on the DFT and the NEGF combined with the two-probe model (electrode-nanostructure-electrode), implemented in ATOMISTIX TOOLKIT (ATK) package,\cite
{17,18,19} which has been widely used to study electron transport through atomic-scale junctions. In this package the electron wave function is expanded by strictly confined basis functions centered at the atoms, radial parts of which are the eigenfunctions of the pseudo-atom within a spherical box with the desired cutoff radius $r_{l}^{c}$, that is, the angular-momentum-dependent numerical eigenfunctions $\varphi_{l}(r)$ of the atomic pseudopotential $V_{l}(r)$ for an energy $\varepsilon_{l}+\delta\varepsilon_{l}$ chosen so that the first node occurs at $r_{l}^{c}$ :\cite
{22}
\begin{eqnarray}
\left(-\frac{1}{2r}\frac{d^{2}}{dr^{2}}r+\frac{l(l+1)}{2r^{2}}+V_{l}(r)\right)\varphi_{l} (r)\nonumber\\
=\left(\varepsilon_{l}+\delta\varepsilon_{l}\right)\varphi_{l} (r),
\end{eqnarray} 
with $\varphi_{l}(r_{l}^c)=0$. When we apply this package to the FE problem, one challenge is how to describe the electron wave function and the electron density in the vacuum between the nanostructure tip and the anode, where there are no atoms and thus there are no basis functions, but there exists electronic current, because the commonly used confinement radii of basis orbitals are 2$\sim$3\AA . To solve this problem, we extend the scope of basis orbitals so that basis set is sufficient to describe the electron density in the vacuum, by decreasing the energy shift $\delta\varepsilon$. In the two-probe system the emitted current can be calculated by the Landauer-B\"uttiker formula\cite
{23}
\begin{eqnarray}
I(V_{b})=\frac{2e}{h}\int T(E,V_{b})[f_{L}(E)-f_{R}(E)]\mathrm{E}
\end{eqnarray} 
where $f_{L}(E)$ and $f_{R}(E)$ represent the Fermi-Dirac distribution functions at the left and right electrodes, respectively, and $T(E,V_{b})$ stands for the transmission coefficient as a function of the electron energy $E$ and bias voltage $V_{b}$.
\begin{figure}
\includegraphics[width=0.33\textwidth]{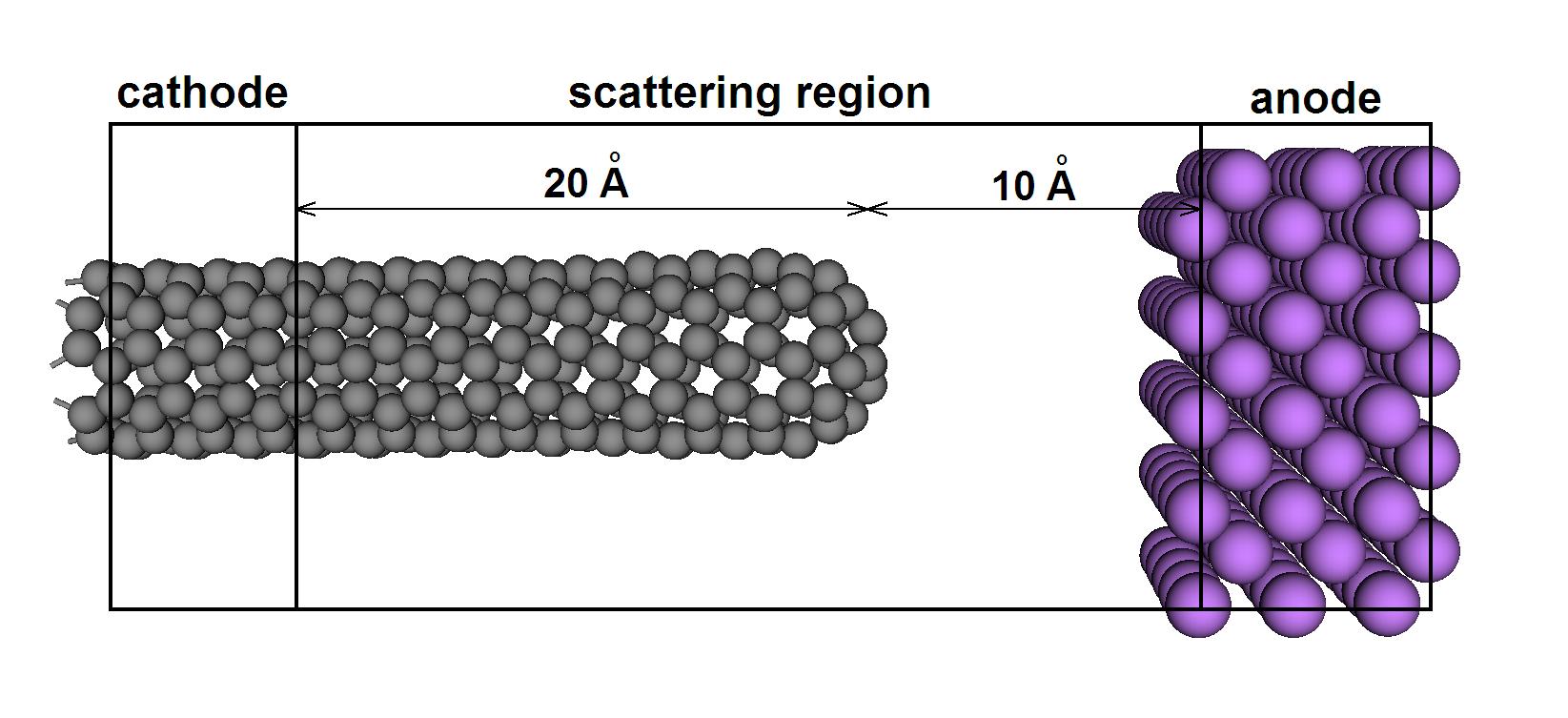}
\caption{The simulated two-probe system. The scattering region consists of 8 unit cells of the (5,5) SWNT capped with a half of a $C_{60}$ molecule and the vacuumwith the length of 10\AA . The cathode and the anode are simulated by 3 unit cells of the SWNT and 6 layers of Li(100) 5x5 surface, respectively.}
\label{fig:1}
\end{figure}
\begin{figure}
\includegraphics[width=0.3\textwidth]{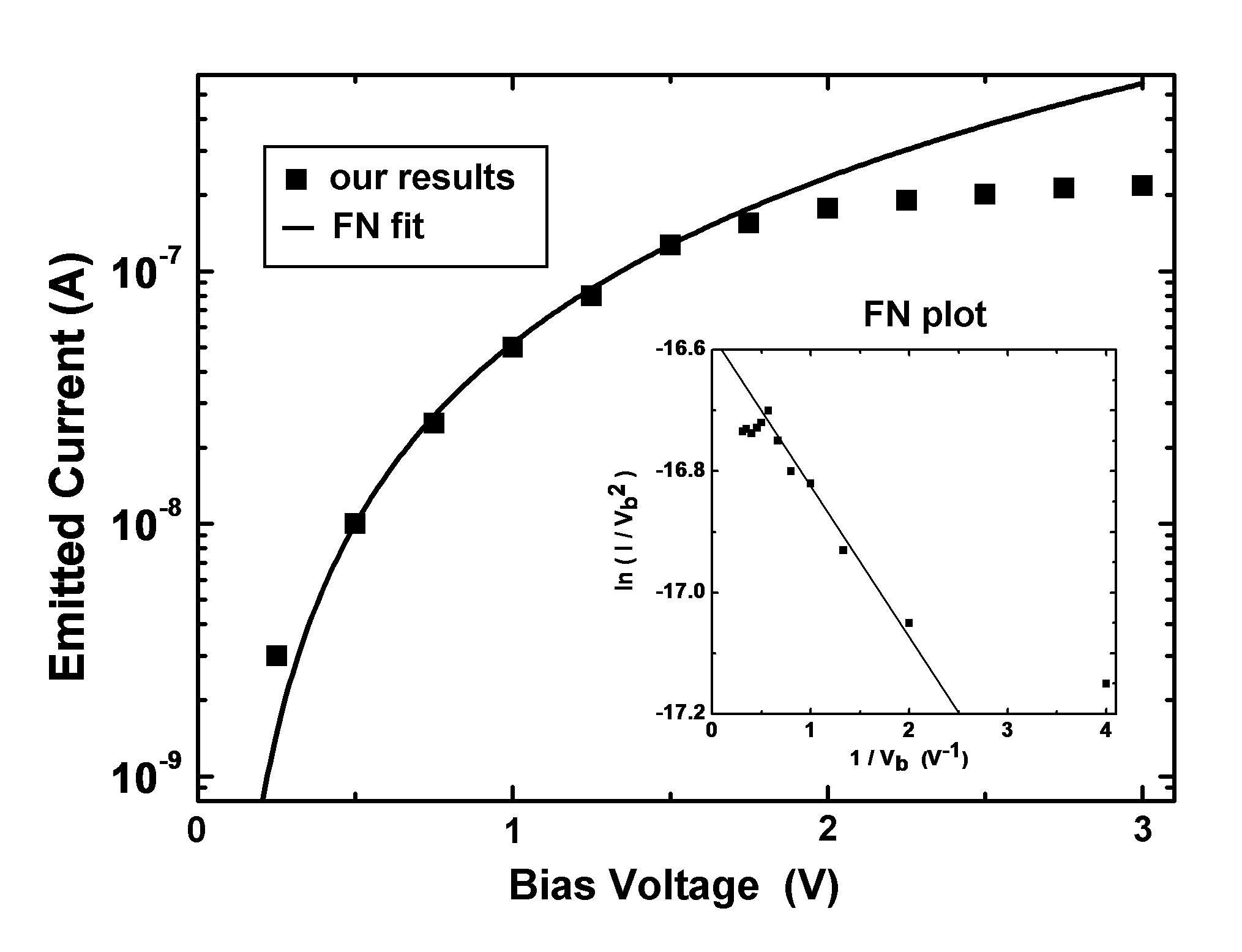}
\caption{The I-V characteristics of the simulated system (black squares) with a FN fit for comparison. Inset figure shows the corresponding plot on FN scales. The current saturation behavior and the deviation from FN fit at high bias are quite apparent.}
\label{fig:2}
\end{figure}
Using the proposed method we have simulated 8 unit cells of a (5,5) SWNT capped with semisphere of a $C_{60}$ molecule, as shown in Fig. 1. The distance between the cathode (left electrode) and the tip is 20\AA which is enough to eliminate effects of the cathode. Computational limitations prevent a longer nanotube to be included. Whether the selected length is adequate can be verified by performing the simulation with various nanotube lengths and determining the length at which the emission current no longer changes for the same electric field. The vacuum size, that is, the distance between the tip and anode (right electrode), was set 10\AA .  Whether this is sufficient can be also checked by examining the gradient of the effective potential in the vacuum: if the local field is higher than the applied electric field then the vacuum level is perturbed by the nanotube, but if the local field is equal to the applied field it means that the anode is sufficiently far from the nanotube. For the cathode of the two-probe system we have selected 3 unit cells of the same (5,5) SWNT, considering that most of the nanotube, except very close to the tip, remains equipotential under the electric field as mentioned above. The anode was simulated by 6 layers of the Li(100) 5x5 surface to separate adjacent nanotubes by at least 10\AA . The lithium electrode has already been utilized in several computational studies.\cite
{24,28} 
 
 Here we note that the lengths of the electrodes have to be large enough, because we would use basis orbitals with large cutoff radii. The system is heterogeneous along the transport direction and thus even at zero bias the system is not periodic in this direction. When calculate the electrostatic potential, we therefore employed a Poisson solver which combined the FFT method in the perpendicular directions with a multigrid solver for the transport direction,\cite
{29}
 where Dirichlet boundary conditions were used for the open system. We have used double zeta plus polarization (DZP) basis set for carbon and lithium atoms and the cutoff radii of the basis orbitals were extended up to 10\AA . Whether the selected basis set is sufficient to achieve convergent results can be verified by performing the calculations with more basis orbitals and larger cutoff radii. The exchange-correlation interaction was described by the Perdew-Wang (PW91) parametrized generalized gradient approximation (GGA)\cite
{30}
 and the electronic temperature was set 300K.
\begin{figure}
\includegraphics[width=0.35\textwidth]{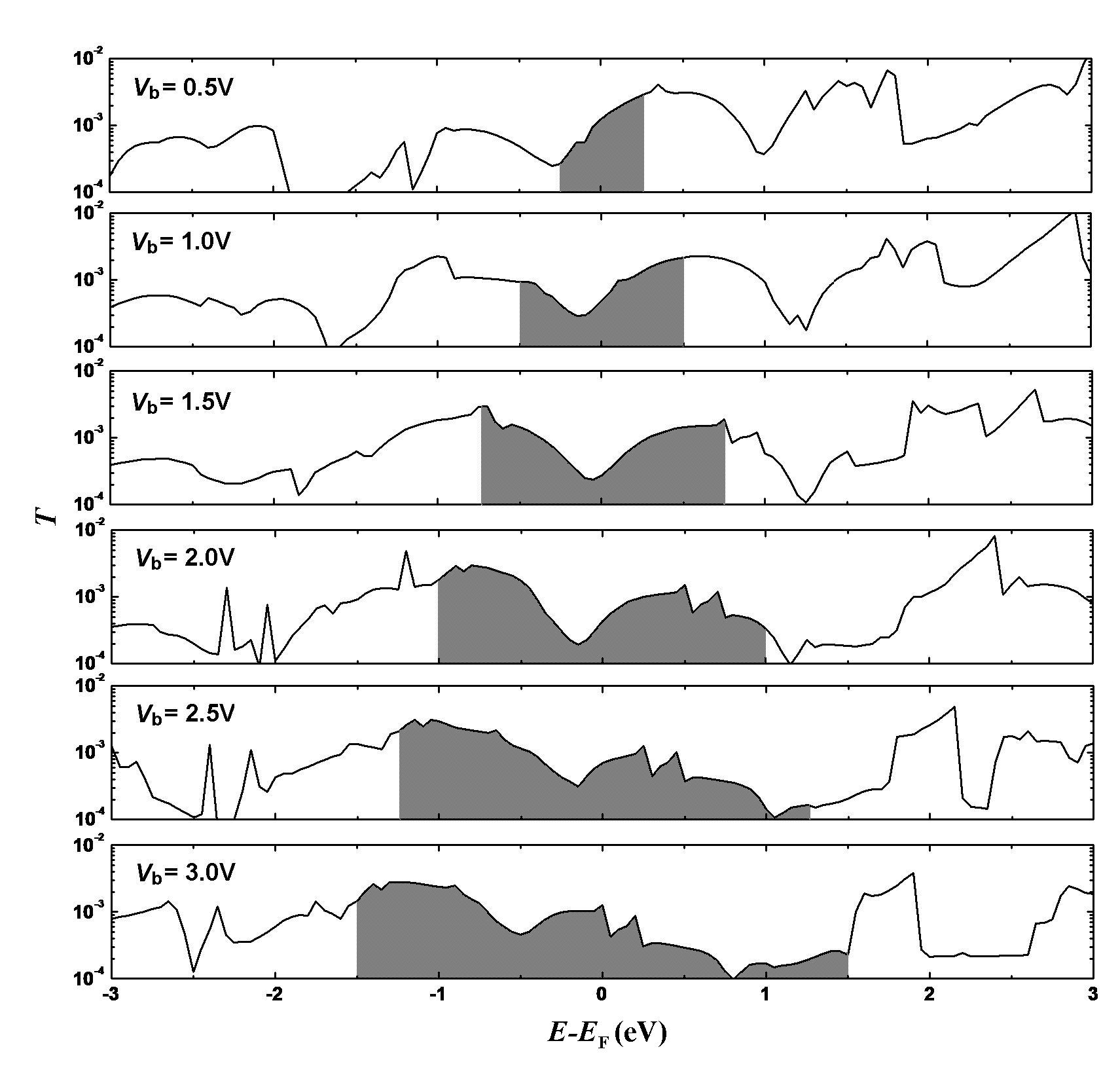}
\caption{Transmission spectrum at various bias voltages. The integrated transmission in the bias window represented by the shaded areas increases with bias voltage increasing at low bias, while remains nearly unchanged at high bias.}
\label{fig:3}
\end{figure}
First, geometry optimization of the capped SWNT was performed such that the force on each atom was less than 0.05eV/\AA . During the relaxation the atoms in the most left 3 unit cells were fixed. Next, using the optimized capped nanotube as input structure, the current-voltage ($I-V$) characteristics were calculated and the results are shown in Fig. 2 with a FN fit for comparison. The corresponding plot on FN scales is also shown in the inset figure, which obviously deviates from a straight line and exhibits a non-FN behavior. The current saturation behavior, which has been previously observed experimentally in single-walled carbon nanotubes,\cite
{31} is quite evident.

Insight can be gained into this saturation behavior by examining the transmission spectrum (Fig. \ref{fig:3}). In the figure shaded areas represents the integrated transmission in the bias window, which is referred to the energy interval from the chemical potential of the left electrode to that of the right electrode. The higher integrated transmissions in the bias window, the higher emitted current. It is clearly seen that the integrated transmission in the bias window increases with bias voltage increasing at low bias, resulting in the pseudoexponential increase of emitted current. But at high bias the integrated transmission remains nearly unchanged, although the bias window gets wider with bias voltage increasing, which explains the current saturation behavior. Our results suggest that the saturation behavior of emitted current may be common properties of nanostructure-based electron emitters, as observed experimentally in single-walled\cite
{31} 
and multi-walled\cite
{32}
 carbon nanotubes, and other sharp emitters like nanowires\cite
{33,34}
 as well.
 
In conclusion, we introduced a general first-principles approach for calculating electron FE characteristics of any kind of nanostructures, based on the DFT combined with the NEGF technique. The method employs basis orbitals with large confinement radii sufficient to describe the electron density in the vacuum between the nanostructure tip and the anode, and takes the presence of emitted current into account. The simulation results on a capped single-walled carbon nanotube reproduce the experimental trend closely, in particular, the current saturation and deviation from the Fowler-Nordheim behavior.

\end{document}